\documentclass[11pt]{article}

\usepackage[T1]{fontenc}
\usepackage[utf8]{inputenc}
\usepackage[
left=2.5cm,
right=2.5cm,
top=2.5cm,
bottom=3cm
]{geometry}
\usepackage{setspace}
\usepackage{cite}
\usepackage{amsmath,amssymb,mathtools}
\usepackage{bm}
\usepackage{mathrsfs}
\usepackage{authblk}
\usepackage{microtype}
\usepackage[hidelinks]{hyperref}

\RequirePackage{xcolor}

\newcommand{\dd}{\mathrm{d}}
\newcommand{\E}{\mathbb{E}}
\newcommand{\Prob}{\mathbb{P}}
\newcommand{\erf}{\operatorname{erf}}
\newcommand{\erfc}{\operatorname{erfc}}
\newcommand{\cO}{\mathcal{O}}
\newcommand{\cI}{\mathcal{I}}

\newcommand{\Om}{\Omega}

\title{\textbf{Distance--Higuchi Bounds on Inflationary Field Ranges and Lifetimes}}

\author[1,2,3]{Luis A. Anchordoqui}
\author[4,5]{Ignatios Antoniadis}
\author[6,7]{Alek Bedroya}

\affil[1]{Department of Physics and Astronomy, Lehman College, City University of New York, NY 10468, USA}
\affil[2]{Department of Physics, Graduate Center, City University of New York, NY 10016, USA}
\affil[3]{Department of Astrophysics, American Museum of Natural History, NY 10024, USA}
\affil[4]{Laboratoire de Physique Th\'eorique et Hautes \'Energies -- LPTHE, Sorbonne Universit\'e, CNRS, 4 Place Jussieu, 75005 Paris, France}
\affil[5]{Center of Excellence in High Energy Physics, , Faculty of Science, Chulalongkorn University, Phayathai Road, Pathumwan,   Bangkok 1030, Thailand}
\affil[6]{Princeton G   ravity Initiative, Princeton, NJ 08544, USA}
\affil[7]{Department of Physics, Princeton University, Princeton, NJ 08544, USA}

\date{July 2026}

\begin{document}

\maketitle

\begin{abstract}
  \noindent We show that quasi-de Sitter inflation driven by a nearly flat scalar potential has a  finite polynomial lifespan dictated by the interplay between the swampland Distance Conjecture and the generalized Higuchi bound. By analyzing classical scalar rolling and quantum stochastic diffusion, we demonstrate that for any arbitrarily high but fixed statistical confidence, a universe cannot live longer than $\sim \min\left\{\ln(\frac{1}{H})\frac{\sqrt V}{V'},V^{-\frac{d-1}{2}}\ln^2(\frac{1}{H})\right\}$ in reduced Planck units  while remaining Higuchi-consistent, where $H$ is the Hubble parameter, $V$ is the scalar potential, $d$ is the number of spacetime dimensions, and prime denotes derivative with respect to the scalar field. This bound despite being weaker than the Trans-Planckian Censorship Conjecture, which has been argued for classical cosmologies that flow to the asymptotic of the field space without tunneling, is powerful given its minimal quantum gravity input and applicability to all points in the moduli space. 
\end{abstract}

\newpage 

\tableofcontents

\section{Introduction}

Understanding the lifetime and stability of quasi-de~Sitter phases remains a central challenge in string phenomenology and early-universe cosmology. Within the Swampland program~\cite{Vafa:2005ui}, the so-called \textit{Swampland conjectures}~\cite{Palti:2019pca,vanBeest:2021lhn,Agmon:2022thq} place severe theoretical constraints on standard cosmological paradigms~\cite{Andriot:2026lac,Anchordoqui:2026hys}. In particular, the Distance Conjecture~\cite{Ooguri:2006in,Etheredge:2022opl}, the de~Sitter Conjecture~\cite{Obied:2018sgi,Ooguri:2018wrx}, and the Trans-Planckian Censorship Conjecture (TCC)~\cite{Bedroya:2019snp,Bedroya:2019tba} tightly constrain the field excursions, potential gradients, and duration of accelerated expansion allowed in an effective theory admitting a consistent ultraviolet completion in quantum gravity. For an approximately constant Hubble parameter $H$, the TCC implies the lifetime bound
\begin{equation}
\tau_{\rm qdS}\lesssim\tau_{\rm TCC}\equiv\frac{1}{H}\ln\left(\frac{M_{\rm pl}}{H}\right),
\end{equation}
where $M_{\rm pl}$ denotes the reduced Planck mass. The conjecture is motivated by its non-trivial realization in weakly coupled and asymptotic regimes of string theory~\cite{Andriot:2020lea,Rudelius:2022gbz,Andriot:2022xjh}, its close quantitative relations to the broader web of Swampland criteria~\cite{Bedroya:2019snp,Brahma:2019vpl, Andriot:2020lea,Bedroya:2020rmd}, independent holographic and effective field theory (EFT)-consistency arguments supporting it in asymptotic regions of scalar-field space or along classical trajectories flowing toward such regions~\cite{Bedroya:2022tbh,vandeHeisteeg:2023uxj,Bedroya:2024zta}, and the longstanding difficulty of constructing parametrically controlled de~Sitter vacua in string theory~\cite{Dine:1985he,Maldacena:2000mw,Andriot:2024cct}. 

Independent of TCC, an extremely conservative upper bound on the life-time of semi-classical $d$-dimensional de Sitter is provided by the de~Sitter Poincaré recurrence time,
\begin{equation}
\tau_{\rm rec}\sim H^{-1}e^{S_{\rm dS}},\qquad {\rm with} \qquad S_{\rm dS}=\frac{4\pi^{(d+1)/2}}{\Gamma\left(\frac{d-1}{2}\right)}\left(\frac{M_{{\rm pl},d}}{H}\right)^{d-2},
\end{equation}
which is exponentially large in the de~Sitter entropy~\cite{Gibbons:1977mu,Dyson:2002pf}. It is therefore natural to ask whether a parametrically shorter, polynomial upper bound can be derived from bottom-up consistency conditions, that unlike~\cite{Bedroya:2022tbh,Bedroya:2024zta}, apply to inflationary cosmologies that settle into a meta-stable de Sitter. This is the aim of the present work. As quantum-gravity input, we assume the emergence of an exponentially light massive spin-2 state over trans-Planckian field distances, which always occurs in known string theory examples and is motivated by the Emergent String Conjecture~\cite{Lee:2019wij} and spin-2 Swampland Conjecture~\cite{Klaewer:2018yxi}. We show that unitarity (i.e. de~Sitter Higuchi bound~\cite{Higuchi:1986py}), together with a quantum-field-theoretic treatment of fluctuations, yields a polynomial upper bound on the lifetime of a quasi-de~Sitter phase.

Moving an infinite distance in field space triggers the exponential descent of a tower of massive states, threatening the validity of the low-energy EFT. While these light fields are usually analyzed via their scalar or spin-1 properties, their higher-spin components, specifically spin-2 states, introduce severe consistency requirements in curved spacetimes. In a de Sitter background, low-mass spin-2 fields are uniquely constrained by the  Higuchi--Deser--Waldron unitarity bound~\cite{Higuchi:1986py,Higuchi:1989gz,Deser:2001us}, below which the helicity-0 mode becomes a ghost, restricting viable inflationary
models across string theory and the Swampland program. For example, in
perturbative string theory settings, this bound dictates the behavior
of Regge trajectories and sets constraints on the allowed string scale~\cite{Lust:2019lmq}. Besides, when analyzed within the Swampland program, the Higuchi lower bound on infinite towers of higher-spin states automatically translates into a strict upper bound on inflaton field excursions, severely limiting large-field displacement~\cite{Scalisi:2019gfv,Antoniadis:2024ypf}. Conversely, the generalized Higuchi bound~\cite{Fasiello:2013woa} enforces an upper limit on the first derivative of the scalar potential, and when combined with the de Sitter Conjecture, this upper limit defines a narrow, stable phenomenological window where slow-roll inflation can survive~\cite{Luben:2020wim}.

Drawing from these compelling observations, in this paper we focus on bounding the lifetime of quasi-dS expansion rather than the field range. First, we map out the finite region of field space where the inflationary EFT remains consistently ghost-free. Then, we analyze this boundary through both classical trajectories and quantum stochastic fluctuations. Classically, the field space restriction translates into a strict lifetime bound on quasi-de Sitter expansion for ordinary slow-roll and exponential tracker potentials. In ultra-flat regimes where classical motion freezes, we employ the framework of stochastic inflation to show that diffusive quantum fluctuations inevitably drive a critical fraction of quantum branches across the ``distance-Higuchi wall.'' Ultimately, by leveraging only the sharpened Distance Conjecture and standard low-energy spectral assumptions, we establish a robust, framework-independent upper limit on the duration of healthy cosmic inflation.

The layout of the paper is as follows. We begin in Sec.~\ref{sec:2} by establishing a field-excursion bound for a scalar field $\phi$ in a $d$-dimensional Friedmann–Lema\^{\i}tre–Robertson–Walker (FLRW) universe, combining the Distance Conjecture with a generalized Higuchi bound. We derive a formula for $\Delta\phi$ based on a generalized stability requirement that avoids spin-2 ghosts, and we provide specific solutions for general and exponential potentials. Armed with our findings, in Sec.~\ref{sec:3} we establish classical upper bounds on the duration of inflation by combining field-range constraints with scalar speed, resulting in bounds that diverge as potentials flatten and the classical slope approaches zero. Our analysis  defines general bounds for shallow slow-roll potentials, highlighting a dependency on the inverse square root of slow-roll parameters, and derives exact, finite lifetime bounds for exponential potential tracker solutions.
After that in Sec.~\ref{sec:4} we model the long-wavelength evolution of a field in an extremely shallow potential as a stochastic process following the perspective advocated in~\cite{Guleryuz:2026rdz}, deriving a finite lifetime bound where quantum diffusion forces an order-one fraction of inflationary branches to cross the distance-Higuchi wall. We demonstrate  this occurs on a timescale scaling as $\mathcal{O}(H^{-(d-1)}\ln^2(1/H))$, providing a conservative limit when classical drift is negligible. We finish with some conclusions presented in Sec.~\ref{sec:5}.

\section{Inflationary field excursion from the Distance Conjecture and the Higuchi bound}
\label{sec:2}

Throughout, we work in $d>2$ spacetime dimensions and use reduced Planck units, so that the coefficient of the Einstein--Hilbert term is one half. We consider a spatially flat FLRW background sourced by one canonically normalized scalar $\phi$
evolving under a potential $V(\phi)$, governed by the action
\begin{equation} S=\int \dd^d x\sqrt{-g}\left[\frac{1}{2}R-\frac{1}{2}(\partial\phi)^2-V(\phi)\right], \label{eq:action} \end{equation}
where $\sqrt{-g}$ denotes the square root of the negative determinant of the FLRW metric tensor $g_{\mu \nu}$, with line element given by
\begin{equation}
\dd s^2=-\dd t^2+a^2(t) \, \dd\bm{x}_{d-1}^{\,2},
\label{FLRW_line_element}
\end{equation}
and where $R$ is the Ricci scalar, $a(t)$ is the scale factor as a function of cosmic time $t$, and $\bm{x}$ denotes the spatial coordinates. The homogeneous equations of motion are
\begin{equation} \frac{(d-1)(d-2)}{2}H^2 =\frac{1}{2}\dot\phi^2+V, \label{eq:friedmann}
\end{equation}
\begin{equation}
  \dot H =-\frac{\dot\phi^2}{d-2}, \label{eq:raychaudhuri}
\end{equation}
and
\begin{equation}
  \ddot\phi+(d-1)H\dot\phi+V' =0 \,, \label{eq:scalarEOM}
\end{equation}
where $H = \dot a/a$ is the Hubble parameter, while the dot and prime denote derivatives with respect to $t$ and $\phi$, respectively. It is useful to introduce
\begin{align} \epsilon_H\equiv-\frac{\dot H}{H^2}=\frac{\dot\phi^2}{(d-2)H^2}. \label{eq:epsilonH} \end{align}
Accelerated expansion is equivalent to $\epsilon_H<1$. We orient the field so that the cosmological solution moves toward increasing $\phi$, and increasing $\phi$ is also the infinite-distance direction in which the tower becomes light. With $\Delta\phi\equiv\phi-\phi_i\geq0$, we assume that the tower contains a massive spin--2 state whose mass obeys
\begin{align} m_2(\Delta\phi)\leq M_{\rm pl}e^{-\alpha \Delta\phi}\,. \label{eq:towermass} \end{align}
The sharpened Distance Conjecture proposes that the lightest tower in an infinite-distance limit has an exponent satisfying~\cite{Ooguri:2006in,Etheredge:2022opl}
\begin{align} \alpha\geq\alpha_\star\equiv\frac{1}{\sqrt{d-2}}. \label{eq:sharpDC} \end{align}
We will use the sharpened Distance Conjecture, however, the exact value of the constant $\alpha$ will not affect our later conclusions. The Distance Conjecture by itself does not specify the spin content of the lightest tower. We therefore make the additional spectral assumption that a spin--2 member of the relevant tower falls at least as rapidly as in \eqref{eq:towermass}. This is automatic for a Kaluza--Klein graviton tower, but it is not a consequence of \eqref{eq:sharpDC} alone.

We first review the argument of Scalisi~\cite{Scalisi:2019gfv}. A totally symmetric massive spin-$s$ field (of mass $m_s$) in exact $d$-dimensional de Sitter space obeys the Higuchi--Deser--Waldron unitarity bound \cite{Higuchi:1986py,Higuchi:1989gz,Deser:2001us}
\begin{align} m_s^2>H^2(s-1)(s+d-4), \qquad s\geq2. \label{eq:spinsHiguchi} \end{align}
At equality the representation becomes partially massless. If $m_s(\Delta\phi)\leq m_{s,0}e^{-\alpha \Delta\phi}$ while $H$ is constant, then the tower mass and the Higuchi lower bound can be compatible only when
\begin{align} \Delta\phi<\frac{1}{\alpha}\ln\left[\frac{M_{\rm pl}}{H\sqrt{(s-1)(s+d-4)}}\right]. \label{eq:ScalisiBound} \end{align}
This is the $d$-dimensional form of Scalisi's field-range bound. The first nontrivial bosonic case is $s=2$,
\begin{align} \Delta\phi<\frac{1}{\alpha}\ln\left(\frac{M_{\rm pl}}{\sqrt{d-2}\,H}\right). \label{eq:ScalisiSpin2} \end{align}
Using the smallest exponent allowed by \eqref{eq:sharpDC} gives the largest conservative interval,
\begin{align} \Delta\phi<L_{\rm dS}(H)\equiv\sqrt{d-2}\ln\left(\frac{M_{\rm pl}}{\sqrt{d-2}\,H}\right). \label{eq:flatWall} \end{align}

The assumption of constant $H$ is unnecessary. In the standard minimally coupled Fierz--Pauli spectator construction on a scalar-supported FLRW background, the helicity-zero Stueckelberg field $\pi$ has a two-derivative term proportional to \cite{Grisa:2009yy,Kolb:2023dzp}
\begin{align} \mathscr{ L}_{\pi,\mathrm{der}} \propto-2m_2^2\left[\frac{d-1}{d-2}m_2^2g^{\mu\nu}-R^{\mu\nu}\right]\nabla_\mu\pi\nabla_\nu\pi \, \label{eq:piAction} \end{align}
where $R^{\mu \nu}$ is the Ricci tensor. For the metric in \eqref{FLRW_line_element},
\begin{align} R^{00}=-(d-1)(H^2+\dot H). \label{eq:R00} \end{align}
The coefficient of $\dot\pi^2$ changes sign when
\begin{align} m_2^2=(d-2)(H^2+\dot H). \label{eq:kineticZero} \end{align}
Consequently, absence of the helicity-zero ghost requires the generalized Higuchi inequality
\begin{align} m_2^2>(d-2)(H^2+\dot H)=(d-2)H^2(1-\epsilon_H). \label{eq:FRWHiguchi} \end{align}
This is exact within the stated quadratic spin--2 EFT and is not a slow-roll expansion. In exact de Sitter it reduces to \eqref{eq:spinsHiguchi} for $s=2$. Away from de Sitter, saturation of \eqref{eq:FRWHiguchi} merely makes the kinetic term vanish; it does not produce the partially massless gauge symmetry. Other curvature couplings can modify the stability condition and the lower bound to other $\mathcal{O}(H^2)$ expressions. However, such modifications will not affect our conclusions.

We now derive a field-range bound for a general non-constant potential. Since $\dot\phi>0$, equations \eqref{eq:raychaudhuri} and \eqref{eq:epsilonH} imply
\begin{align} \dot\phi=H\sqrt{(d-2)\epsilon_H}, \qquad {\rm or \ equivalently} \qquad \frac{\dd\ln H}{\dd\phi}=-\sqrt{\frac{\epsilon_H}{d-2}}. \label{eq:HamiltonJacobi} \end{align}
Let the trajectory begin at $(\phi_i,H_i)$ and end at $\phi_f=\phi_i+\Delta\phi$. Integrating \eqref{eq:HamiltonJacobi} gives
\begin{align} H_f=H_i\exp\left[-\int_0^{\Delta\phi}\dd \Delta\phi\sqrt{\frac{\epsilon_H(\Delta\phi)}{d-2}}\right]. \label{eq:HfGeneral} \end{align}
At the endpoint, \eqref{eq:towermass} and \eqref{eq:FRWHiguchi} require
\begin{align} M_{\rm pl}e^{-\alpha\Delta\phi}>H_f\sqrt{(d-2)(1-\epsilon_f)}, \qquad {\rm with} \qquad \epsilon_f\equiv\epsilon_H(\phi_f). \label{eq:endpointConsistency} \end{align}
Eliminating $H_f$ with \eqref{eq:HfGeneral} yields the exact integrated constraint
\begin{align} \int_0^{\Delta\phi}\dd \Delta\phi\left[\alpha-\sqrt{\frac{\epsilon_H(\Delta\phi)}{d-2}}\right]<{\cal B}, \qquad {\rm where} \qquad {\cal B}\equiv\ln\left[\frac{M_{\rm pl}}{H_i\sqrt{(d-2)(1-\epsilon_f)}}\right]. \label{eq:generalIntegratedBound} \end{align}
The physical content of this equation is simple. The tower mass falls as $e^{-\alpha \Delta\phi}$, while the generalized Higuchi scale falls because $H$ decreases. The two effects compete through the difference between $\alpha$ and $\sqrt{\epsilon_H/(d-2)}$. For an accelerated solution, $\epsilon_H<1$, and the sharpened value $\alpha_\star$ is always larger than the rate at which $H$ decreases.

If $0\leq\epsilon_H(\Delta\phi)\leq\epsilon_{\max}<1$ on the interval, then \eqref{eq:sharpDC} and \eqref{eq:generalIntegratedBound} imply
\begin{align} \Delta\phi<\frac{\sqrt{d-2}}{1-\sqrt{\epsilon_{\max}}}\,{\cal B}. \label{eq:generalFieldBound} \end{align}
This formula is a convenient generalization of \eqref{eq:flatWall}. It reduces to the de Sitter result as $\epsilon_{\max}\to0$ and becomes weak as the expansion approaches the non-accelerating boundary $\epsilon_H=1$.

The result is especially transparent for an exponential potential
\begin{align} V(\phi)=V_i e^{-\lambda(\phi-\phi_i)}, \qquad {\rm with} \qquad  0<\lambda<\frac{2}{\sqrt{d-2}}. \label{eq:exponentialPotential} \end{align}
The scalar-dominated scaling solution has
\begin{align} \epsilon_\lambda=\frac{(d-2)\lambda^2}{4}, \qquad {\rm and} \qquad H(\Delta\phi)=H_i e^{-\lambda \Delta\phi/2}. \label{eq:expHphi} \end{align}
The inequality on $\lambda$ in \eqref{eq:exponentialPotential} is precisely $\epsilon_\lambda<1$, namely the condition for accelerated expansion. 
Combining the exponential decrease of the tower mass with the exponential decrease of $H$ gives
\begin{align} (\alpha-\lambda/2)\Delta\phi<{\cal B}_\lambda \,,\label{eq:expIntermediate} \end{align}
where
\begin{align}
  {\cal B}_\lambda\equiv\ln\left(\frac{M_{\rm pl}}{C_\lambda H_i}\right) \qquad {\rm with} \qquad C_\lambda\equiv\sqrt{(d-2)(1-\epsilon_\lambda)}. \label{eq:Blambda}
\end{align}
All in all, the exact field-excursion bound on the scaling solution is
\begin{align} \Delta\phi<\frac{{\cal B}_\lambda}{\alpha-\lambda/2}. \label{eq:expFieldGeneralAlpha} \end{align}
Using the sharpened Distance Conjecture coefficient gives the weakest universal form
\begin{align} \boxed{\Delta\phi<\Delta\phi_{\max}^{\exp}\equiv\frac{\sqrt{d-2}}{1-\lambda\sqrt{d-2}/2}\ln\left[\frac{M_{\rm pl}}{H_i\sqrt{(d-2)\left(1-(d-2)\lambda^2/4\right)}}\right].} \label{eq:expFieldSharp} \end{align}
The denominator is positive exactly in the accelerated range $\lambda<2/\sqrt{d-2}$. The flat-potential result \eqref{eq:flatWall} follows smoothly as $\lambda\to0$. The bound weakens as $\lambda$ approaches the critical value because $H$ then decreases almost as rapidly as the slowest tower allowed by the sharpened Distance Conjecture, while the no-ghost threshold in \eqref{eq:FRWHiguchi} also tends to zero as $\epsilon_\lambda\to1$.

\section{Classical upper bounds on the duration of inflation}
\label{sec:3}

The field-range bounds become lifetime bounds once they are combined with the classical speed of the scalar. Equation \eqref{eq:epsilonH} gives the exact identity
\begin{align} T=\int_{\phi_i}^{\phi_f}\frac{\dd\phi}{H(\phi)\sqrt{(d-2)\epsilon_H(\phi)}}. \label{eq:exactTimeIntegral} \end{align}
A finite classical upper bound therefore requires a nonzero lower bound on the classical slope, or equivalently on $\epsilon_H$. This is why the classical argument becomes weak for an extremely flat potential.

Next, for a general monotonic inflationary interval, we define
\begin{align} H_{\min}\equiv\min_{\phi_i\leq\phi\leq\phi_f}H(\phi), \qquad \epsilon_{\min}\equiv\min_{\phi_i\leq\phi\leq\phi_f}\epsilon_H(\phi), \qquad {\rm and} \qquad \epsilon_{\max}\equiv\max_{\phi_i\leq\phi\leq\phi_f}\epsilon_H(\phi). \label{eq:extrema} \end{align}
Bearing this in mind and assuming $\epsilon_{\min}>0$ and $\epsilon_{\max}<1$, equations \eqref{eq:exactTimeIntegral} and \eqref{eq:generalFieldBound} imply
\begin{align} T<\frac{{\cal B}}{H_{\min}\sqrt{\epsilon_{\min}}\left(1-\sqrt{\epsilon_{\max}}\right)}. \label{eq:generalClassicalTime} \end{align}
This form keeps the variation of the background explicit. It is conservative because it uses the smallest allowed tower exponent, the smallest Hubble scale, and the smallest classical speed over the whole interval.

For a shallow potential, the usual slow-roll equations in $d$ dimensions are
\begin{align} H^2\simeq\frac{2V}{(d-1)(d-2)}, \qquad {\rm and} \qquad (d-1)H\dot\phi\simeq-V'. \label{eq:slowrollEquations} \end{align}
It is convenient to define
\begin{align} \epsilon_V\equiv\frac{d-2}{4}\left(\frac{V'}{V}\right)^2, \qquad {\rm and} \qquad \eta_V\equiv\frac{d-2}{2}\frac{V''}{V}. \label{eq:potentialSlowRoll} \end{align}
Then $\epsilon_H\simeq\epsilon_V$ and
\begin{align} |\dot\phi|\simeq H\sqrt{(d-2)\epsilon_V}. \label{eq:slowrollSpeed} \end{align}
If $H$ and $\epsilon_V$ vary only by subleading fractions over the allowed excursion, then \eqref{eq:flatWall} and \eqref{eq:slowrollSpeed} lead to 
\begin{align} \boxed{T_{\rm cl}^{\rm shallow}\lesssim\frac{1}{H\sqrt{\epsilon_V}}\ln\left(\frac{M_{\rm pl}}{\sqrt{d-2}\,H}\right).} \label{eq:shallowClassicalTime} \end{align}
The associated number of e-folds satisfies
\begin{align} N_{\rm cl}^{\rm shallow}\lesssim\frac{1}{\sqrt{\epsilon_V}}\ln\left(\frac{M_{\rm pl}}{\sqrt{d-2}\,H}\right). \label{eq:shallowClassicalN} \end{align}
Equation \eqref{eq:generalClassicalTime} supplies the controlled version when the slow-roll parameters vary appreciably. The important parametric feature is the factor $\epsilon_V^{-1/2}$. The classical upper bound diverges in the exactly flat limit because the homogeneous solution then does not move through field space.\footnote{This is the so-called ultra-slow roll inflationary regime where the second slow-roll parameter is large $\eta=-3$ leading to an enhancement of the power spectrum due to a growing mode~\cite{Tsamis:2003px, Kinney:2005vj}.} On the other hand, it gives a theoretical upper limit on the scale of inflation $H$, or equivalently on the ratio $r$ of tensor-to-scalar primordial perturbations when it applies to single field slow-roll inflation models in $d=4$:
\begin{align}  
N\lesssim 4\pi\sqrt{{\cal A}_s}\frac{M_\text{pl}}{\sqrt{2}H}\ln\frac{M_\text{pl}}{\sqrt{2}H}\,,
\label{eq:Nlimit4d} 
\end{align}
where ${\cal A}_s =H^2/(8\pi^2\epsilon_V)=2.1\times 10^{-9}$ is the amplitude of the power spectrum and $r=16\epsilon_V$, evaluated at the horizon exit. Using the approximate bound $N\gtrsim \ln(M_I/{\rm eV})$, where $M_I\simeq\sqrt{H M_{\rm pl}}$ is the inflationary scale, and combining it with \eqref{eq:Nlimit4d}, we obtain
\begin{align}
H\lesssim 1.6\times10^{14}\,{\rm GeV}\simeq6.5\times10^{-5}M_{\rm pl} \qquad {\rm and} \qquad r\lesssim0.41\,,
\label{eq:BoundsHr}
\end{align}
which is weaker than the 95\% CL experimental bound on $r$ by roughly an order of magnitude ($r < 0.032$, derived using a combination of BICEP/Keck 2018 and Planck data~\cite{BICEP:2021xfz,Tristram:2021tvh}).

We next obtain an exact classical lifetime bound for the exponential potential \eqref{eq:exponentialPotential}. The scalar-dominated scaling, or power-law, solution is \cite{Lucchin:1984yf,Halliwell:1986ja}
\begin{align} a(t)=a_i\left(\frac{t}{t_i}\right)^p, \qquad \phi(t)=\phi_i+\frac{2}{\lambda}\ln\left(\frac{t}{t_i}\right), \qquad p=\frac{4}{(d-2)\lambda^2}. \label{eq:trackerSolution} \end{align}
To verify the solution, the exponential potential must scale as $t^{-2}$, which fixes the coefficient of $\ln t$ in $\phi$ to $2/\lambda$. Equation \eqref{eq:raychaudhuri} then gives $p=4/[(d-2)\lambda^2]$, and \eqref{eq:scalarEOM} fixes
\begin{align} V_i=\frac{2[(d-1)p-1]}{\lambda^2t_i^2}  \qquad {\rm and} \qquad H_i=\frac{p}{t_i}. \label{eq:trackerNormalization} \end{align}
The solution has
\begin{align} \epsilon_H=\frac{1}{p}=\epsilon_\lambda=\frac{(d-2)\lambda^2}{4}. \label{eq:trackerEpsilon} \end{align}
It inflates precisely when $p>1$, which reproduces the range in \eqref{eq:exponentialPotential}. Taking $t_f=t_i+T$, the field excursion is
\begin{align} \Delta\phi=\frac{2}{\lambda}\ln\left(\frac{t_f}{t_i}\right)=\frac{2}{\lambda}\ln\left(1+\frac{H_iT}{p}\right). \label{eq:trackerExcursionTime} \end{align}
Combining this relation with \eqref{eq:expFieldGeneralAlpha} gives
\begin{align} \ln\left(\frac{t_f}{t_i}\right)<\frac{\lambda}{2\alpha-\lambda}{\cal B}_\lambda. \label{eq:trackerRatioBound} \end{align}
Therefore, for a general distance exponent $\alpha>\lambda/2$,
\begin{align} T<\frac{p}{H_i}\left[\left(\frac{M_{\rm pl}}{C_\lambda H_i}\right)^{\lambda/(2\alpha-\lambda)}-1\right]. \label{eq:trackerTimeGeneralAlpha} \end{align}
Using $\alpha=\alpha_\star$ gives the following bound
\begin{align} \boxed{T_{\rm cl}^{\exp}\lesssim\frac{1}{\lambda^2 H_i}\left[\left(\frac{M_{\rm pl}}{H_i\sqrt{1-(d-2)\lambda^2/4}}\right)^{\lambda\sqrt{d-2}/\left(2-\lambda\sqrt{d-2}\right)}\right].} \label{eq:trackerTimeSharp} \end{align}
There are $\mathcal{O}(1)$ factors which we did not include as they are bounded both from below and above. These constants depend on the initial mass of the tower as well as the exact generalization of the Higuchi bound based on the assumptions made about the interactions of the massive spin-2 particle in the tower. It is still helpful to have a concrete inequality for specific set of assumptions. Under the generalizations we used for the Higuchi bound and assuming the initial mass of the tower is less than $M_{\rm pl}$, restoring the numerical coefficient leads to the following inequality.
\begin{align} T_{\rm cl}^{\exp}<\frac{4}{(d-2)\lambda^2H_i}\left[\left(\frac{M_{\rm pl}}{H_i\sqrt{(d-2)\left(1-(d-2)\lambda^2/4\right)}}\right)^{\lambda\sqrt{d-2}/\left(2-\lambda\sqrt{d-2}\right)}-1\right].  \end{align}
The number of e-folds is $N=p\ln(t_f/t_i)$ and obeys
\begin{align} N_{\rm cl}^{\exp}<\frac{2}{\lambda\sqrt{d-2}\left(1-\lambda\sqrt{d-2}/2\right)}\ln\left[\frac{M_{\rm pl}}{H_i\sqrt{(d-2)\left(1-(d-2)\lambda^2/4\right)}}\right].
\label{eq:trackerNSharp} \end{align}

Along the exponential tracker, the field displacement per e-fold is constant,
\begin{equation}
\frac{\dd\phi}{\dd N}=\frac{(d-2)\lambda}{2} 
\qquad\Longrightarrow\qquad
N=\frac{2\Delta\phi}{(d-2)\lambda}.
\end{equation}
When inflation reaches the Higuchi wall, the endpoint condition is saturated,
\begin{equation}
M_{\rm pl} e^{-\Delta\phi_f/\sqrt{d-2}}
=
H_f\sqrt{(d-2)\left(1-\frac{(d-2)\lambda^2}{4}\right)}.
\end{equation}
Solving for $\Delta\phi_f$ and substituting into the tracker relation therefore gives
\begin{align}
\boxed{
N_{\rm cl}^{\exp}
<
\frac{2}{\lambda\sqrt{d-2}}
\ln\left[
\frac{M_{\rm pl}}
{H_f\sqrt{(d-2)\left(1-(d-2)\lambda^2/4\right)}}
\right].
}
\label{eq:trackerNSharpHf}
\end{align}

The shallow limit of the exact tracker result agrees with the slow-roll estimate. Defining $\lambda\sqrt{d-2}/2=\sqrt{\epsilon_\lambda}$, for $\epsilon_\lambda\ll1$,
\begin{align} T_{\rm cl}^{\exp}=\frac{{\cal B}_\lambda}{H_i\sqrt{\epsilon_\lambda}}\left[1+\cO(\sqrt{\epsilon_\lambda}\,{\cal B}_\lambda)\right], \label{eq:trackerShallowLimit} \end{align}
which is \eqref{eq:shallowClassicalTime}. This also makes clear why the classical argument alone becomes arbitrarily weak as the potential is flattened.

\section{Quantum diffusion and the lifetime of very shallow potentials}
\label{sec:4}

For an extremely shallow potential, the homogeneous drift can be too small for \eqref{eq:shallowClassicalTime} to be useful. The long-wavelength scalar nevertheless evolves because modes continuously cross the Hubble scale. A useful heuristic argument starts from the de Sitter thermal estimate $\langle\dot\phi^2\rangle\sim H^d$. This does not imply coherent motion with speed $H^{d/2}$ for the entire duration of inflation. Hubble friction erases the velocity memory on a time of order $H^{-1}$. The displacement during one correlated interval is therefore $\delta\phi\sim H^{d/2}H^{-1}=H^{(d-2)/2}$. During a time $T$ there are $n\sim HT$ independent steps, whose signs add randomly rather than coherently. Hence
\begin{align} \Delta\phi_{\rm rms}\sim\sqrt{n}\,\delta\phi\sim\sqrt{H^{d-1}T}. \label{eq:randomWalkHeuristic} \end{align}
Reaching a wall a distance $L$ away therefore takes a typical time $T\sim L^2/H^{d-1}$. Since the distance--Higuchi wall has $L\sim\ln(1/H)$, this already explains the scaling $H^{-(d-1)}\ln^2(1/H)$.

We now derive the coefficient and formulate the statement precisely. In exact de Sitter space, we use conformal time $a(\eta)=-1/(H\eta)$. A free massless canonical scalar has Bunch--Davies modes
\begin{align} u_k(\eta)=\frac{\sqrt{\pi}}{2}e^{i\vartheta_d}H^{(d-2)/2}(-\eta)^{(d-1)/2}\text{H}^{(1)}_{(d-1)/2}(-k\eta), \label{eq:BDmode} \end{align}
where $\text{H}^{(1)}_\nu$ is the Hankel function of the first kind and the phase is irrelevant. On super-Hubble scales $k\eta\to 0$,
\begin{align} |u_k|^2\longrightarrow\frac{2^{d-3}}{\pi}\Gamma^2\left(\frac{d-1}{2}\right)H^{d-2}k^{-(d-1)}. \label{eq:lateMode} \end{align}
Using the surface area of a $(d-2)$-dimensional unit sphere embedded in a $(d-1)$-dimensional space,
\begin{align} \Om_{d-2}=\frac{2\pi^{(d-1)/2}}{\Gamma\left(\frac{d-1}{2}\right)}, \label{eq:sphereArea} \end{align}
the field variance per logarithmic momentum interval is
\begin{align} {\cal P}_\phi(k)\equiv\frac{\Om_{d-2}}{(2\pi)^{d-1}}k^{d-1}|u_k|^2=A_dH^{d-2}, \qquad {\rm where} \qquad 
A_d\equiv\frac{\Gamma\left(\frac{d-1}{2}\right)}{2\pi^{(d+1)/2}}. \label{eq:AdDefinition} \end{align}
The average field displacement $\delta\phi$ can be deduced by the 2-point function at equal points in the position space, which can be computed upon momentum integration with appropriate treatment of ultraviolet and infrared divergences~\cite{Ratra:1984yq,Antoniadis:2020bwi}:
\begin{align} 
(\delta\phi)^2\equiv \langle\phi^2\rangle = A_dH^{d-2}N + c_d + {\cal O}(e^{-N})\,,
\label{eq:phi2} 
\end{align}
where we have switched to the e-fold number $N=HT$ and expanded at large $N$; $c_d$ is an irrelevant order 1 constant (for instance $c_4=1/2$). The unitarity Higuchi bound wall is at distance
\begin{align} L_\star=\sqrt{d-2}\ln\left(\frac{M_{\rm pl}}{\sqrt{d-2}\,H}\right)\,, \label{eq:stochasticWall} \end{align}
imposing $(\delta\phi)^2<L_\star^2$ and thus
\begin{align} 
N_q\lesssim 
\label{eq:boundNq} \frac{d-2}{A_d}H^{-(d-2)}\ln^2\left(\frac{M_{\rm pl}}{\sqrt{d-2}H}\right)\,,
\end{align}
where the subindex $q$ denotes its quantum origin.

An alternative method to obtain the bound is based on the stochastic approach. We coarse-grain the field at $k_c=\sigma_c aH$, with fixed $\sigma_c\ll1$. During $\dd t$, the shell entering the long-wavelength sector has $\dd\ln k_c=H\dd t$ at leading order. The infinitesimal increase in the variance of the long-wavelength field $\phi_L$,
\begin{align} \dd\E[(\phi_L-\E\phi_L)^2]=A_dH^{d-1}\dd t \,,
\label{eq:varianceRate} \end{align}
quantifies how much more spread out the value of the $\phi_L$ becomes over the time interval $\dd t$ as new random quantum modes cross the horizon and lock into place. Note that $\E$ denotes the expectation value. Independent crossing shells generate Gaussian noise. After the decaying momentum mode has relaxed, the long-wavelength stochastic EFT is 
\begin{align} \dd \Delta\phi=b(\Delta\phi)\dd t+\sqrt{A_dH^{d-1}}\,\dd W_t, \qquad {\rm with} \qquad b(\Delta\phi)=-\frac{V'(\phi_i+\Delta\phi)}{(d-1)H}, \label{eq:Langevin} \end{align}
where $\Delta\phi \equiv \phi_L - \phi_i$ is the stochastic fluctuation (or displacement) of the long-wavelength scalar field away from its initial background value, and $W_t$ is a standard Wiener process (Brownian motion), with first and second moments given by
\begin{align} \E[\dd W_t]=0, \qquad {\rm and} \qquad \E[\dd W_t\dd W_{t'}]=\delta(t-t')\dd t\,\dd t', \label{eq:Wiener} \end{align}
respectively~\cite{Starobinsky:1994bd,Vennin:2015hra,Pattison:2019hef}. The first moment shows that random fluctuations have a mean of zero; on average, the noise does not push the field in any specific direction. When evaluated at identical times ($t = t'$), the second moment equation reduces to the standard It\^o  calculus property $\E[(\dd W_t)^2] = \dd t$. Therefore, the expected value of the squared increment matches the time increment itself.

The corresponding Fokker--Planck equation is found to be
\begin{align} \frac{\partial P}{\partial t}=-\frac{\partial}{\partial \Delta\phi}(bP)+\frac{A_dH^{d-1}}{2}\frac{\partial^2P}{\partial \Delta\phi^2} \,, \label{eq:FokkerPlanck} \end{align}
where $P$ is the probability density function  of the stochastic variable $\Delta\phi$ at time $t$. For the orientation chosen above, a potential that rolls toward the light-tower direction has $V'<0$ and hence $b\geq0$. Such a drift can only make the wall be reached earlier. The zero-drift problem therefore gives a conservative upper bound on the time for which an order-one fraction of branches remains healthy.

Set $b=0$ and take $H$ to be constant at leading quasi-de Sitter order. Then
\begin{align} \Delta\phi_t=\sqrt{A_dH^{d-1}}\,W_t, \qquad {\rm and} \qquad \E[\Delta\phi_t^2]=A_dH^{d-1}t. \label{eq:Brownian} \end{align}
Let $L$ be the actual distance--Higuchi wall and define the first crossing time
\begin{align} \tau_L\equiv\inf\{t\geq0:\Delta\phi_t\geq L\}. \label{eq:hittingTime} \end{align}
The reflection principle gives the exact first-passage probability \cite{Ando:2020fjm}
\begin{align} \Prob(\tau_L\leq T)=\erfc\left(\frac{L}{\sqrt{2A_dH^{d-1}T}}\right), \qquad {\rm or \ else} \qquad  \Prob(\tau_L>T)=\erf\left(\frac{L}{\sqrt{2A_dH^{d-1}T}}\right). \label{eq:firstPassage} \end{align}
For each stochastic branch, let $\cI_T=1$ if the branch has remained on the healthy side of the wall up to time $T$, and let $\cI_T=0$ otherwise. Then
\begin{align} \E[\cI_T]=\Prob(\tau_L>T). \label{eq:indicatorExpectation} \end{align}
Thus, $\E[\cI_T]$ is the ensemble-averaged healthy fraction. Moving forward, we fix a number $0<\delta<1$, independent of $H$, and demand that no more than a fraction $\delta$ of branches violate the Higuchi condition:
\begin{align} \Prob(\tau_L\leq T)\leq\delta, \qquad {\rm or \  equivalently} \qquad \E[\cI_T]\geq1-\delta. \label{eq:fixedFraction} \end{align}

In the very shallow limit, the actual wall is no farther away than $L_\star$ given in \eqref{eq:stochasticWall}.
Since the crossing probability decreases as the wall is moved outward,
\begin{align} \Prob(\tau_L\leq T)\geq\erfc\left(\frac{L_\star}{\sqrt{2A_dH^{d-1}T}}\right)\,, \label{eq:probabilityLowerBound} \end{align}
where $\erfc=1-\erf$ and $\erf$ is the error function.
The requirement \eqref{eq:fixedFraction} can therefore hold only if
\begin{align} T\leq\frac{L_\star^2}{2A_dH^{d-1}[\erfc^{-1}(\delta)]^2}. \label{eq:stochasticIntermediate} \end{align}
Substitution of \eqref{eq:AdDefinition} and \eqref{eq:stochasticWall} gives the fixed-fraction quantum lifetime bound
\begin{align} \boxed{T_{\rm q}^{(\delta)}\leq\frac{(d-2)\pi^{(d+1)/2}}{\Gamma\left(\frac{d-1}{2}\right)[\erfc^{-1}(\delta)]^2}H^{-(d-1)}\ln^2\left(\frac{M_{\rm pl}}{\sqrt{d-2}\,H}\right)} \,, \label{eq:quantumLifetime} \end{align}
which is essentially the same with \eqref{eq:boundNq} using $N=HT$.
For every fixed order-one $\delta$, this quantum lifetime and the corresponding number of e-folds scale as
\begin{align} T_{\rm q}^{(\delta)}=\cO\left(H^{-(d-1)}\ln^2\frac{1}{H}\right) \qquad {\rm and} \qquad N_{\rm q}^{(\delta)}=\cO\left(H^{-(d-2)}\ln^2\frac{1}{H}\right), \label{eq:quantumScaling} \end{align}
respectively. A non-negative classical drift toward the wall only strengthens \eqref{eq:quantumLifetime}. Indeed, when the drift is approximately constant, the drifted and driftless processes can be coupled to the same Wiener path as $\Delta\phi_t^{(b)}=\Delta\phi_t^{(0)}+bt$, so $\tau_L^{(b)}\leq\tau_L^{(0)}$ path by path.

The result also admits a comparison form for a slowly varying quasi-de Sitter geometry. Suppose that during the candidate interval $H(t)\geq H_->0$ and $\epsilon_H(t)\leq\epsilon_+<1$, and that the drift is directed toward the tower. The local distance--Higuchi wall is then no farther than
\begin{align} L_+\equiv\sqrt{d-2}\ln\left[\frac{M_{\rm pl}}{H_-\sqrt{(d-2)(1-\epsilon_+)}}\right]. \label{eq:comparisonWall} \end{align}
The quadratic variation of the stochastic part obeys
\begin{align} {\cal A}(T)=\int_0^T A_dH^{d-1}(t)\dd t\geq A_dH_-^{d-1}T. \label{eq:quadraticVariation} \end{align}
The Brownian time-change theorem and the reflection principle therefore imply
\begin{align} \Prob(\tau_{\rm loss}\leq T)\geq\erfc\left(\frac{L_+}{\sqrt{2A_dH_-^{d-1}T}}\right). \label{eq:comparisonProbability} \end{align}
Consequently,
\begin{align} T\leq\frac{(d-2)\pi^{(d+1)/2}}{\Gamma\left(\frac{d-1}{2}\right)[\erfc^{-1}(\delta)]^2}H_-^{-(d-1)}\ln^2\left[\frac{M_{\rm pl}}{H_-\sqrt{(d-2)(1-\epsilon_+)}}\right]. \label{eq:comparisonLifetime} \end{align}
This form does not require an expansion of the generalized Higuchi bound in $\epsilon_H$.

The finite quantity in \eqref{eq:quantumLifetime} is a fixed quantile, or equivalently the time at which the ensemble-averaged healthy fraction falls below a specified order-one value. It is not the literal mean first-passage time. The one-sided driftless density is
\begin{align} f_L(t)=\frac{L}{\sqrt{2\pi A_dH^{d-1}t^3}}\exp\left[-\frac{L^2}{2A_dH^{d-1}t}\right]. \label{eq:LevyDensity} \end{align}
Since $f_L(t)\sim t^{-3/2}$ at late times,
\begin{align} \E[\tau_L]=\int_0^\infty t f_L(t)\dd t=\infty. \label{eq:meanDiverges} \end{align}
There is also no finite deterministic time by which every branch has crossed. The physically meaningful statement is that an order-one fraction crosses by the finite time \eqref{eq:quantumLifetime}. A finite mean exit time would require an additional loss boundary, a drift toward the wall, or another mechanism that removes the long one-sided survival tail.

The classical and quantum bounds exchange dominance at an extremely small slope. At leading shallow-potential order,
\begin{align} T_{\rm cl}^{\rm shallow}\sim\frac{B_0}{H\sqrt{\epsilon_V}} \qquad {\rm and} \qquad T_{\rm q}^{(\delta)}\sim K_{d,\delta}H^{-(d-1)}B_0^2 \,, \label{eq:twoScales} \end{align}
where
\begin{align} K_{d,\delta}\equiv\frac{(d-2)\pi^{(d+1)/2}}{\Gamma\left(\frac{d-1}{2}\right)[\erfc^{-1}(\delta)]^2} \qquad {\rm and} \qquad  B_0\equiv\ln\left(\frac{M_{\rm pl}}{\sqrt{d-2}\,H}\right). \label{eq:Kdefinition} \end{align}
The stochastic bound is stronger when
\begin{align} \sqrt{\epsilon_V}\lesssim\frac{H^{d-2}}{K_{d,\delta}B_0}. \label{eq:crossover} \end{align}
Thus, quantum diffusion does not usually improve the classical bound at an ordinary slow-roll slope. Its role is precisely to close the loophole in which the potential is so flat that the classical traversal time would otherwise become arbitrarily large.

\section{Conclusions}
\label{sec:5}

Our main result is a bottom-up bound on the region of field space in which an inflationary EFT can remain healthy. The basic mechanism follows from the interplay between the Distance Conjecture and the Higuchi bound. A massive spin--2 state in the lightest tower associated with the inflationary field has a mass that decreases exponentially with the scalar displacement until it eventually violates the generalized Higuchi bound and its helicity-zero mode becomes ghostlike. This defines a finite distance--Higuchi region within which the inflationary EFT can remain consistent.

For a sufficiently shallow potential, as relevant for slow-roll inflation, the resulting field-range constraint takes the form
\begin{equation}
\Delta\phi\lesssim \sqrt{d-2}\ln\left(\frac{M_{\rm pl}}{H}\right),
\end{equation}
up to order-one factors inside the logarithm. Thus, even when the potential is arbitrarily shallow, a quasi-de Sitter EFT cannot remain ghost-free over an arbitrarily large scalar excursion toward the direction in which the spin--2 tower becomes light.

The corresponding bounds on the duration of inflation follow from the time required for the scalar to explore this finite field range. For ordinary slow-roll evolution, classical drift gives
\begin{equation}
\tau_{\rm cl}\lesssim \ln\left(\frac{1}{H}\right)\frac{\sqrt V}{|V'|},
\end{equation}
where we work in reduced Planck units and suppress dimension-dependent order-one coefficients. Equivalently, $\tau_{\rm cl}\sim H^{-1}\epsilon_V^{-1/2}\ln(1/H)$. This bound becomes weak in the limit $V'\rightarrow0$, since the homogeneous classical solution then moves arbitrarily slowly through field space.

Quantum fluctuations provide an independent constraint precisely in this ultra-flat regime. Although the characteristic velocity satisfies $\langle\dot\phi^2\rangle\sim H^d$, the velocity is correlated only over a Hubble time. Successive Hubble-time displacements therefore add diffusively which combined with the distance--Higuchi field range yields
\begin{equation}
\tau_{\rm q}\lesssim H^{-(d-1)}\ln^2\left(\frac{1}{H}\right).
\end{equation}
or equivalently, the number of e-folds are bounded as
\begin{align}
    N_{\rm q}\lesssim H^{-(d-2)}\ln^2\left(\frac{1}{H}\right)\,.
\end{align}
Whenever both descriptions are applicable, the duration of the inflationary phase is consequently bounded by
\begin{equation}
\boxed{\tau_{\rm inf}\lesssim\min\left\{\ln\left(\frac{1}{H}\right)\frac{\sqrt V}{|V'|},V^{-\frac{d-1}{2}}\ln^2\left(\frac{1}{H}\right)\right\}.}
\end{equation}

The classical constraint controls ordinary slow-roll evolution, whereas the stochastic constraint closes the limit in which the classical traversal time diverges as the potential becomes arbitrarily flat.

The stochastic bound requires a slightly more careful interpretation. It is not an absolute time by which every stochastic branch must cross the distance--Higuchi wall. Instead, for any fixed $0<\delta<1$, it bounds the time for which the fraction of branches that have crossed the wall can remain smaller than $\delta$. More precisely,
\begin{equation}
T_{\rm q}^{(\delta)}\lesssim\frac{H^{-(d-1)}}{[\operatorname{erfc}^{-1}(\delta)]^2}\ln^2\left(\frac{1}{H}\right),
\end{equation}
up to dimension-dependent coefficients. The choice of $\delta$ therefore modifies only an order-one prefactor. For every fixed $\delta$, including values arbitrarily close to unity, the parametric dependence $T_{\rm q}^{(\delta)}\sim H^{-(d-1)}\ln^2(1/H)$ remains unchanged. This result concerns the ensemble fraction of coarse-grained stochastic branches, or equivalently the descendants of an initially specified comoving Hubble patch.

The quantum-gravity assumptions entering the argument are deliberately limited. We assume the sharpened Distance Conjecture~\cite{Etheredge:2022opl} and more specifically the existence of a massive spin--2 state whose mass decreases along the relevant scalar direction as $m_2\lesssim m_0e^{-\alpha\Delta\phi}$, with $\alpha\geq 1/\sqrt{d-2}$. In the ultra-flat regime we additionally use the standard stochastic description of the long-wavelength scalar. Subject to these assumptions, a quasi-de Sitter region cannot remain arbitrarily long-lived while simultaneously remaining Higuchi-consistent along the scalar direction in which the spin--2 tower becomes exponentially light. Our analysis makes precise the expectation that the Distance Conjecture should constrain inflation through the eventual breakdown of the low-energy description. 

It is useful to compare these constraints with the much stronger consequences of the TCC. In four-dimensional single-field slow-roll
 inflationary cosmologies that settle into a meta-stable de Sitter, assuming that the observed scalar perturbations are generated by the inflaton and imposing the minimum duration required to account for the observable universe, the classical bound gives approximately
\begin{equation}
H \lesssim1.6\times10^{14}\,{\rm GeV}\simeq6.5\times10^{-5}M_{\rm pl} \qquad {\rm and} \qquad r\lesssim0.41.
\end{equation}
The corresponding inflationary energy scale can therefore remain of order $10^{16}~{\rm GeV}$. The stochastic bound is considerably weaker for ordinary slow-roll slopes and becomes relevant only in the ultra-flat regime. Our bounds are thus substantially less restrictive than those obtained from the TCC~\cite{Bedroya:2019tba}, but they follow from a small set of quantum-gravity assumptions and unlike some of the other bottom-up contraints~\cite{Bedroya:2022tbh,Bedroya:2024zta}, apply locally in field space. 

In summary, armed with the sharpened Distance Conjecture and the Higuchi bound we have demonstrated that a quasi-de Sitter region cannot be arbitrarily long-lived while remaining Higuchi-consistent on an order-one fraction of its coarse-grained quantum branches. Although this bound is weaker than the TCC, which applies to classical cosmologies moving toward the boundaries of field space without tunneling, it remains highly effective. Its strength lies in requiring minimal quantum gravity input and being applicable to all points in the moduli space.

\section*{Acknowledgements}

L.A.A. is supported by the U.S. National Science Foundation
(NSF Grant PHY-2412679), he extends his appreciation to the Harvard Swampland Initiative for their warm hospitality and for providing a stimulating environment for productive discussions. I.A. is supported by the Second Century Fund (C2F), Chulalongkorn University and in part by the Higher Education and Science Committee of MESCS RA (Research Project N 24RL-1C036); he also thanks the hospitality and support of the Institute for Advanced Study in Princeton, where this work was initiated. A.B. is supported in part by the Simons Foundation under grant number 654561 and by the Princeton Gravity Initiative at Princeton University.

\end{document}